\documentclass[12pt]{article}
\usepackage[a4paper, margin=2cm]{geometry}
\usepackage[utf8]{inputenc}
\usepackage[T1]{fontenc}
\usepackage{hyperref}
\usepackage{amsmath}
\usepackage{amsfonts}
\usepackage{amssymb}
\usepackage{amsthm}
\usepackage{amsbsy}
\usepackage{mathrsfs}
\usepackage{slashed}
\usepackage{graphicx}
\usepackage{tikz-cd}
\usepackage{fancyhdr}

\numberwithin{equation}{section}

\newtheoremstyle{henrique}      %Name
  { }                                              % Space above
  {}                                              % Space below
  {\normalfont}                            % Body font
  {}                                              % Indent amount
  {\bfseries}                                % Theorem head font
  {.}                                             % Punctuation after theorem head
  { }                                             % Space after theorem head, ' ', or \newline
  {}                                              % Theorem head spec (can be left empty, meaning `normal')
\theoremstyle{henrique}

\hypersetup{colorlinks=false, urlbordercolor={1 0 0}, linkbordercolor={0 0 1}, citebordercolor={0 0 1}}

\usepackage{cite}

\newcommand{\da}{\dot{A}}
\newcommand{\db}{\dot{B}}
\newcommand{\dc}{\dot{C}}
\newcommand{\ep}{\epsilon}
\newcommand{\dD}{\dot{D}}
\newcommand{\de}{\dot{E}}
\newcommand{\dM}{\dot{M}}
\newcommand{\dN}{\dot{N}}

\newcommand{\dd}{\mathrm{d}}
\newcommand{\vv}{\mathrm{v}}

\title{Non-local conformal symmetry in Fronsdal theory.}
\author{Henrique Flores\thanks{henrique@ift.unesp.br}.}
\date{}

\begin{document}

\begin{titlepage} 
\pagenumbering{gobble}
\maketitle 
\vspace{-1.0cm}
\begin{center}
\small{\textit{Instituto de F\'{i}sica Te\'{o}rica, Universidade Estadual Paulista, \\
Rua Dr. Bento Teobaldo Ferraz 271, \\
Bloco II -- Barra Funda, \\
CEP: 01140-070 -- S\~{a}o Paulo, Brasil.}}
\end{center}
\vspace{.1cm}
\begin{abstract}
We write a first order action for higher-spin fields and
construct a canonical map to Fronsdal theory. The first-order description is defined over complex field configurations and
has conformal invariance. We show that it is possible to push forward these transformations 
to a set of symmetries in Fronsdal theory that satisfies the conformal algebra but is not given
by standard conformal change of coordinates.
\end{abstract}
\tableofcontents
\end{titlepage}

\pagenumbering{arabic}

\section{Introduction.}

Higher-spin theories have an important application in the AdS/CFT context.
When the AdS radius is small, it is conjectured that 
a subset of large string excitations decouples from the remaining degrees of freedom 
and is described by an interacting higher-spin theory\cite{schwa}. 
Unfortunately, interactions are subtle to construct,
but there is a comparatively easier case which we can study:
in the $N \rightarrow \infty$ limit, we have free massless higher-spin theories.

There are two known descriptions of free theories, 
which are referred to as Fronsdal and Penrose formulation. 
In Fronsdal theory, we have constrained spacetime tensors that form an irreducible representation of the little group on-shell,
while in Penrose theory one uses twistor geometry to construct
irreducible representations of the little group.
Both theories are well described by an action which is invariant under higher-spin gauge symmetries.
It is interesting, however, that Penrose formulation \textit{is} invariant under conformal symmetries
while Fronsdal formulation \textit{is not} \cite{Afr}.

At first glance, it may seem strange that two descriptions of the same theory have different symmetries.
The present paper aims to solve this conflict. After a brief review of the two formulations,
we construct the action for Penrose theory
and show that both theories describe the same classical phase space via a canonical transformation. 
Using the canonical map, we can push forward the conformal transformations of Penrose theory
to a set of non-local conformal symmetries in Fronsdal description.

\subsection{Plan of this paper.}

We organize our presentation as follows.
Section $2$ is a brief review,
where we explain the two approaches for free massless higher-spin theories.

In section $3$ we write an action for Penrose higher-spin theory. To our knowledge, such action
for general higher-spins has never appeared before in the literature. A similar action, however, was used
to describe full self-dual gravity in \cite{krasnov:2016}. In our case, this action is defined 
over complex field configurations, and it describes off-shell a doubled set of the higher-spin modes.
In phase space, however, there is a well-defined notion of reality,
and it is where we obtain a single copy of the spectrum. 

It is instructive, at this point, to look at some examples, so
the spins $1$, $3/2$ and $2$ cases are discussed in detail, each of which
highlights a particular feature of our construction
outlining our strategy for dealing with general spins.
The spin $s$ case is done in section $4$. 
We construct the map which relates Fronsdal and Penrose descriptions
and show that both theories describe
the same phase space by mapping one symplectic structure into the other.

With this map, we can investigate conformal invariance. 
In section $5$ we show that Penrose action does have conformal symmetry for every spin $s$.
Therefore one is able to push forward these transformations to the Fronsdal case.
For spins lower than $2$, these new transformations agree with usual conformal change
of coordinates. The first non-trivial case is linearized gravity.
We write explicitly the resulting transformation, where one is able to see the difference from standard Lie derivatives.

\subsection{On notation.}

We are concerned with $4$-dimensional Minskowski space; so, through out the paper, 
the various indices will always be running over fixed intervals.
Small Latin letters, for example, are spacetime indices running from 
$0$ to $3$, so that $A_{m}$ is a spacetime covector.
Capital Latin letters, in turn, are spinor indices in Van der Warden notation, that is, dotted and undotted running from $0$ to $1$.
In particular, a Dirac spinor is a two component Weyl and anti-Weyl spinor written like

\begin{equation}
\Psi = 
\begin{pmatrix}
\psi_{A} \\
\overline{\chi}^{\da}
\end{pmatrix}
\end{equation}

\noindent
for some chiral spinor $\psi_{A}$ and anti-chiral $\overline{\chi}^{\da}$.

Such notation is designed so that there is a correspondence between spacetime and spinor indices
where, for instance, $m$ will correspond to the pair $M \dM$.
The explicit realization is given by the Pauli matrices with
index structure $\sigma^{m}_{M \dM}$, where 

\[
\sigma^{0} = -1 \quad  \text{ and } \quad  \vec{\sigma} = (\sigma^{1}, \sigma^{2} , \sigma^{3}) .
\]

\noindent
The epsilon symbol satisfies $\ep_{AB} \ep^{BC} = \delta_{A}^{\,\,\,\,\,\,C}$ for undotted and
dotted indices. This enables one to raise the indices of $\sigma^{m}_{M \dM}$ to obtain

\[
\overline{\sigma}^{m \, M \dM}, \quad \text{ where } \quad \sigma^{0} = -1 \quad  \text{ and } \quad  \vec{\sigma} = (- \sigma^{1}, -\sigma^{2} ,- \sigma^{3}) .
\]

\noindent
Everything is combined to form the Weyl representation of the Dirac matrices:

\begin{equation}
\Gamma^{m} = 
\begin{pmatrix} 
0  & \sigma^{m} \\
\overline{\sigma}^{m} & 0
\end{pmatrix},
\end{equation}

\noindent
which satisfy the Clifford algebra

\begin{equation}
\left\{ \Gamma^{m}, \Gamma^{n} \right\} = - 2 \eta^{mn}
\end{equation}

\noindent
for the metric signature $(-,+,+,+)$. Our conventions follow those of \cite{wessbagger:susy}.

%For the rest of this paper, of special interest, are symmetrization and anti-symmetrization among indices. We define
%for an arbitrary tensor with $p$ indices, for example $t_{m_{1} \, \cdots \, m_{p}}$,
%the following convention for symmetrization:
%\begin{equation}
%t_{( m_{1} \, \cdots \, m_{p})} = \frac{1}{p!} \sum_{\texttt{over permutations}} t_{m_{i_{1}}  \cdots \, m_{i_{p}}}
%\end{equation}
%\noindent
%while antisymmetry is denoted
%\begin{equation}
%t_{[m_{1} \, \cdots \, m_{p}]} = \frac{1}{p!} \sum_{\texttt{over permutations}} (-)^{\ell(i)} t_{m_{i_{1}}  \cdots \, m_{i_{p}}}
%\end{equation}
%\noindent
%where $\ell$ is the length of the permutation.

\section{Review of massless higher-spin formulations.}

This section is an overview of some background material based on references \cite{fronsdal:boson} and \cite{fangfronsdal:fermion}.
It begins with Fronsdal theory and then proceeds
to Penrose description\cite{penroseRindler:spinortwistor}.
%In its first part, we describe basic aspects of Fronsdal formulation;
%then we proceed to a presentation of Penrose description.
%Massless higher-spins are described by irreducible representations of $so(2)$.	
%On-shell they are characterized by one-degree of freedom called helicity.

\subsection{Fronsdal theory of free massless higher-spin fields.} 
\label{sectionfronsdal}

Let us begin with bosonic spins. Given a totally symmetric tensor of $s$ indices, $h_{m_{1} \cdots m_{s}}$, 
which has higher-spin gauge freedom of the form

\begin{equation}
\delta h_{m_{1} \cdots \, m_{s}} = s \, \partial_{(m_{1}} \varepsilon_{m_{2} \cdots \, m_{s})}
\label{gauge}
\end{equation}

\noindent
and is double-traceless:

\begin{equation}
\eta^{m_{1} m_{2}} \eta^{m_{3} m_{4}} h_{m_{1} m_{2} m_{3} m_{4} \cdots \, m_{s}} = 0;
\label{trace}
\end{equation}

\noindent
one can form the so-called
Fronsdal tensor:

\begin{equation}
F_{m_{1} \cdots \, m_{s}} = \square h_{m_{1} \cdots \, m_{s}} - s \, \partial_{ (m_{1}} \partial^{p} h_{ | p | \, m_{2} \cdots \, m_{s})} + 
\frac{s(s-1)}{2}\partial_{(m_{1}} \partial_{m_{2}} h^{p}_{\,\,\,\,| p | \, m_{3} \cdots \, m_{s})}.
\label{fronsdaltensor}
\end{equation}

\noindent
A higher-spin theory in flat spacetime is then described by the action

\begin{equation}
S = \frac{(-1)}{2}^{s+1} \int \dd^{4} x \left( h^{m_{1} \cdots \, m_{s}} F_{m_{1} \cdots \, m_{s}} - \frac{s(s-1)}{4} h_{n}^{\,\,\,\,n m_{3} \cdots \, m_{s}} F^{p}_{\,\,\,\, p m_{3} \cdots \, m_{s}} \right),
\label{fronsdalaction}
\end{equation}

\noindent
which is symmetric in the higher-spin field $h_{m_{1} \cdots \, m_{s}}$
and gauge invariant under transformations \eqref{gauge}.

The equations of motion read 

\begin{equation}
F_{m_{1} \cdots \, m_{s}} - \frac{s(s-1)}{4}  \eta_{(m_{1} m_{2}} F^{p}_{\,\,\,\, p \, m_{3} \cdots \, m_{s})} =0.
\label{motion}
\end{equation}

\noindent
And these can be further simplified if \eqref{trace} is taken into account. It implies

\begin{equation}
\eta^{m_{1} m_{2}} \eta^{m_{3} m_{4}} F_{m_{1} m_{2} m_{3} m_{4} \cdots \, m_{s}} = 0
\label{b}
\end{equation}

\noindent
which, in turn, allows us to cast equation \eqref{motion} as

\begin{equation}
F_{m_{1} \cdots \, m_{s}} = 0.
\label{fronsdal}
\end{equation}

\noindent
We see the Fronsdal tensor fixes $h_{m_{1} \cdots m_{s}}$ up to gauge transformations 
since both have the same number of degrees of freedom.
The physical degrees of freedom, however, are obtained once we gauge fix the above description.
It is possible to gauge away the trace part of the higher-spin field $h_{m_{1} \cdots m_{s}}$ as well as 
its divergence.
Consider the gauge field $\varepsilon$ which satisfies

\begin{equation}
h^{p}_{\,\,\,\,p m_{3} \cdots \, m_{s}} = \partial^{n} \varepsilon_{n m_{3} \cdots \, m_{s}}
\end{equation}

\noindent
and 

\begin{equation}
\partial^{p} h_{p m_{2} \cdots \, m_{s} } = \square \varepsilon_{m_{2} \cdots \, m_{s} },
\end{equation}

\noindent
so that the remaining gauge symmetry obeys

\begin{equation}
\square \varepsilon_{m_{2} \cdots \, m_{s} } = 0, \quad \partial^{n} \varepsilon_{n m_{3} \cdots \, m_{s}} = 0, \, \text{ and } \, \varepsilon^{p}_{\,\,\,\,p m_{3} \cdots \, m_{s} } = 0.
\label{choices}
\end{equation}

%\vspace{.5cm}
%\noindent
%\textbf{Remark.} The tracelessness of $\varepsilon$ comes from requiring the equations of motion to be divergenceless.

%\vspace{.5cm}
\noindent
Once we choose \eqref{choices}, our higher-spin field satisfies

\begin{equation}
\square h_{m_{1} \cdots \, m_{s}} = 0, \quad  \partial^{p} h_{p m_{2} \cdots \, m_{s}} = 0, \, \text{ and } \, h^{p}_{\,\,\,\, p m_{3} \cdots m_{s}} = 0;
\end{equation}

\noindent
thus proving that $h_{m_{1} \cdots \, m_{s}}$ describes a spin $s$ massless particle.
%\noindent
%thus proving that $h_{m_{1} \cdots \, m_{s}}$ is a $O(2)$ representation.

There are minor changes if one wants to describe fermions.
For a spin $s = h + 1/2$,
we have a Majorana spinor $\Psi_{m_{1} \cdots m_{h}}$ totally symmetric in its $h$ indices
which has gauge freedom

\begin{equation}
\delta \Psi_{m_{1} \cdots m_{h}} = h \, \partial_{(m_{1}} \chi_{m_{2} \cdots m_{h})},
\end{equation}

\noindent
and satisfies the triple $\Gamma$-trace condition: 

\begin{equation}
\Gamma^{m_{1}} \Gamma^{m_{2}} \Gamma^{m_{3}} \Psi_{m_{1} m_{2} m_{3} \cdots m_{h}} = 0.
\label{triple}
\end{equation}

\noindent
The fermionic Fronsdal tensor,

\begin{equation}
F_{m_{1} \cdots m_{h}} = \Gamma^{a} \partial_{a} \Psi_{m_{1} \cdots m_{h}} - h \,  \partial_{(m_{1}} \Gamma^{a} \Psi_{m_{2} \cdots m_{h})a},
\end{equation}

\noindent
is the gauge invariant object used to construct the action

\begin{equation}
S = \frac{1}{2} \int \dd^{4} x  \left( \overline{\Psi}^{m_{1} \cdots m_{h}} 
F_{m_{1} \cdots m_{h}} 
- \frac{h}{2} \, \Gamma^{p}\overline{\Psi}_{p}^{\,\,\,\,\, m_{2} \cdots m_{h}}  \Gamma^{a} F_{a m_{2} \cdots m_{h}} 
- \frac{h(h-1)}{4} \overline{\Psi}_{q}^{\,\,\,\,\,q m_{3} \cdots m_{h}} F^{p}_{\,\,\,\,\,p m_{3} \cdots m_{h}} \right)
\end{equation}

\noindent
where $\overline{\Psi}^{m_{1} \cdots m_{h}}$ satisfies the Majorana condition:

\begin{equation}
\overline{\Psi}^{m_{1} \cdots m_{h}} = \Psi^{T} C, \quad \text{ and } \quad
C = 
\begin{pmatrix}
\ep_{BA} & 0 \\
0 & \ep^{\db\da}
\end{pmatrix}
\label{sec1eq9}
\end{equation}

\noindent
is the charge conjugation matrix.
The equations of motion are

\begin{equation}
F_{m_{1} \cdots m_{s}} - \frac{h}{2} \Gamma_{(m_{1}} \Gamma^{a} F_{m_{2} \cdots m_{s}) a} - \frac{h(h-1)}{4} \eta_{(m_{1} m_{2}} F^{p}_{\,\,\,\,\, m_{3} \cdots m_{s})p} = 0.
\label{eomfermion}
\end{equation}

\noindent
and they can be simplified once one notices \eqref{triple} implies

\begin{equation}
\Gamma^{m_{1}} \Gamma^{m_{2}} \Gamma^{m_{3}} F_{m_{1} m_{2} m_{3} \cdots m_{h}} = 0,
\end{equation}

\noindent
which enables one to cast \eqref{eomfermion} in the form

\begin{equation}
F_{m_{1} \cdots \, m_{h}} = 0.
\end{equation}

\noindent
Notice that, again, the fermionic Fronsdal tensor fixes $\Psi_{m_{1} \cdots m_{h}}$ up to gauge transformations.
The physical degrees of freedom are obtained
from the gauge parameter $\chi_{m_{2} \cdots m_{h}}$ that satisfies

\begin{equation}
\Gamma^{p} \Psi_{p m_{2} \cdots m_{h}} = \Gamma^{m} \partial_{m} \chi_{m_{2} \cdots m_{h}},
\label{fixed}
\end{equation}

\noindent
so that the remaining gauge symmetry obeys

\begin{equation}
\Gamma^{m} \partial_{m} \chi_{m_{2} \cdots m_{h}} = 0 \quad \text{ and } \quad \Gamma^{p} \chi_{p m_{2} \cdots m_{h}} = 0.
\end{equation}

%\vspace{.5cm}
%\noindent
%\textbf{Remark.} The parameter $\chi$ is annihilated by $\Gamma^{m}$ so that the equations of motion are divergenceless.

%\vspace{.5cm}
\noindent
The gauge fixing \eqref{fixed} ensures that $\Psi_{m_{1} \cdots m_{h}}$ is an irreducible
representation of the little group.
The on-shell degrees of freedom are then described by a field $\Psi$ which satisfies

\begin{equation}
\Gamma^{p} \partial_{p} \Psi_{m_{1} \cdots m_{h}} = 0 \quad \text{ and } \quad \Gamma^{p} \Psi_{p m_{2} \cdots m_{h}} = 0
\end{equation}

\noindent
thus proving $\Psi_{m_{1} \cdots m_{h}}$ describes an spin $s=h + 1/2$ representation.

\subsection{Penrose theory of free massless higher-spin fields.} \label{penrose}

Penrose's description of massless higher-spin fields is obtained from the Penrose transform.
It relates homogeneous functions of definite degree in twistor space to
massless higher-spin fields in Minkowski space. For an introduction to twistors,
see reference \cite{tod} as well as references therein.

Here we describe the integral expressions obtained by Penrose in \cite{penroseRindler:spinortwistor} only to give some context.
These integral formulas are not necessary for the rest of this paper. We are only interested in the spacetime fields they define.

Let $Z = ( \omega^{A}, \pi_{\da} )$ be the coordinates of a twistor inside the complex projective line $\mathbf{P}_{1}$.
These are constrained by the twistor equation:

\begin{equation}
\omega^{A} =  x^{A \da} \pi_{\da},
\end{equation}

\noindent
where $x^{A \da}$ parametrizes the Minkowski space. Consider also a point $ \overline{Z} = ( \lambda_{A}, \mu^{\da})$ in the dual twistor space
and fix two closed cycles of integration: $\gamma$ inside $\mathbf{P}_{1}$ and $\gamma^{*}$ inside the dual line $\mathbf{P}_{1}^{*}$.
Define the following spacetime spinors 

\begin{subequations}
\begin{equation}
\overline{\phi}_{\da \db \cdots \, \dD} (x) = \frac{1}{2 \pi i} \int_{\gamma} \underbrace{\pi_{\da} \pi_{\db} \ldots \pi_{\dD}}_{2s} f( Z ) \, \pi_{\de} \dd \pi^{\de}
\label{integral}
\end{equation}

\noindent
and

\begin{equation}
\phi_{A B \cdots \, D} (x) = \frac{1}{ 2 \pi i} \int_{\gamma^{*}} \underbrace{\lambda_{A} \lambda_{B} \ldots \lambda_{D}}_{2s} \overline{f}( \overline{Z} ) \, \lambda^{A} \dd \lambda_{A}
\end{equation}
\label{formulae}
\end{subequations}

\noindent
for some semi-integer number $s$. 

\vspace{.5cm}
\textbf{Remark.} These integrals are well defined over $\mathbf{P}_{1}$
if the integrands are homogeneous functions of degree $0$.
Hence, the complex functions $f( Z )$ and $ \overline{f} (\overline{Z} )$ must have homogeneity $-2s - 2$
in $\pi_{\da}$ and $\lambda_{A}$ respectively.

\vspace{.5cm}
These spinors form an irreducible representation of the Lorentz group
$SL(2, \mathbf{C})$ and satisfy, by consequence of their definitions,
the differential equations 

\begin{equation}
\partial^{A \da} \overline{\phi}_{ \da \db \cdots \dD} (x) = 0
\label{bianchi}
\end{equation}

\noindent
and 

\begin{equation}
\partial^{\da A} \phi_{A B \cdots \, D} (x) = 0.
\label{bianchi2}
\end{equation}

\noindent
%Both equations \eqref{bianchi} and \eqref{bianchi2} 
%restrict $\phi^{AB \cdots D}$ and $\overline{\phi}^{\da \db \cdots \dD}$ to a single component each, one being the complex 
%conjugate of the other.
In view of the (anti-)self-duality conditions, we can see
$\phi^{AB \cdots D}$ and $\overline{\phi}^{\da \db \cdots \dD}$
describe right-handed massless free fields of spin $s$ and left-handed massless free fields of spin $-s$
respectively.

%It is possible to prove that every massless spin $s$ can be obtained from equations \eqref{formulae} via \u{C}ech cohomology \cite{eastwoodPenrose:cohomologys}.
%In such twistor cohomology computations is useful to define a gauge potential. 
Let $a_{\da B \cdots \, D}$ be the field given by 

\begin{equation}
\overline{\phi}_{\da \db \cdots \, \dD} = \partial^{B}_{\,\,\,\,\,\,(\db} \cdots \partial^{D}_{\,\,\,\,\,\,\dD} \, a_{\da) B \cdots \, D}.
\label{definition}
\end{equation}

\noindent
It readily follows that equation \eqref{bianchi} is automatically satisfied when

\begin{equation}
\partial_{(A}^{\,\,\,\,\,\,\da} a_{B \cdots \, D) \da} = 0.
\label{eomgauge}
\end{equation}

\noindent
Notice, however, that there is an ambiguity. 
There are gauge symmetries of the form

\begin{equation}
\delta a_{\da B \cdots \, D} = \partial_{\da (B} \xi_{ \cdots \, D)}
\label{higher}
\end{equation}

\noindent
for some symmetric spinor $\xi_{C \cdots \, D}$ of $2s-2$ indices.
These are the higher-spin gauge symmetries which were also present in Fronsdal theory.

We will always refer to $\phi^{AB \cdots D}$ and $a_{\da B \cdots D}$ as
the fundamental fields of Penrose description. And, for future reference, we call  $\phi_{AB \cdots \, D}$ the curvature spinor
and $a_{\da B \cdots \, D}$ the gauge field.

\section{Higher-spin action in Penrose's description.}

%Fix a real slice inside
%the complexified Minkowski space and let $x^{A \da}$ parametrize this surface. 

\subsection{Higher-spin action.}

We suggest the following higher-spin action for a massless spin $s$ particle:

\begin{equation}
S = i \int \dd^{4} x \, \left( \phi^{AB \cdots \, D} \partial_{A \da} a^{\da}_{\,\,\,\,\,\, B \cdots \, D} \right)
\label{paction}
\end{equation}

\noindent
where $\phi^{AB \, \cdots \, D}$ and $a^{\da}_{\,\,\,\,\, B \cdots \, D}$ have 
$2s$ and $2s-1$ undotted indices respectively.
Invariance under higher-spin gauge symmetries is respected,
because if we consider the variation under \eqref{higher} the action transforms into

\begin{equation}
\delta S = i \int \dd^{4} x \, \left[ \phi^{AB \cdots \, D} \partial_{A \da} \partial^{\da}_{\,\,\,\,\,(B} \xi_{\cdots \, D)} \right].
\end{equation}

\noindent
From the identity 

\begin{equation}
\partial_{A \da} \partial^{\da}_{\,\,\,\,\,B} = + \frac{1}{2} \ep_{AB} \square,
\end{equation}

\noindent
we get $\delta S = 0 $ since the curvature spinor $\phi^{AB \cdots \, D}$ is completely symmetric in its indices.
The equations of motion obtained from \eqref{paction} are precisely \eqref{bianchi} and \eqref{eomgauge}:

\[
\partial_{\da A} \phi^{AB \cdots \, D} = 0 \quad \text{and} \quad \partial_{\da (A} a^{\da}_{\,\,\,\,\,\,B \cdots \, D)} = 0.
\]

\subsection{Reality conditions.}
Although twistors were used as a motivation for this action, we are not integrating over twistor space.
We are only using a spinor basis and it is possible to write this action with usual Lorentz indices too.
The convenience of using spinors is the easier treatment of self-duality conditions. 

A possibly troublesome point is that it appears that this action describes just one helicity, but this is not the case.
Let us discuss this point in detail. 
For the sake of argument, let us specialize our discussion to the spin 1 case.
We want to show that the phase space spanned by these equations is equivalent to the phase
space of Maxwell's electromagnetism.
The natural route is to describe a canonical map.
Therefore, given the data $(\phi, a)$, we are supposed to construct a map
to the Maxwell gauge field $A$,

\begin{equation}
H: \left( \phi, a \right) \longmapsto A,
\end{equation}

\noindent
where solutions of the $(\phi, a)$ system are carried to solutions of the Maxwell's equations. 
In addition, we must verify two things: 
the kernel of this map must be zero, otherwise there are configurations of $\phi$ and $a$ which would 
correspond to zero electromagnetic solution; and the cokernel should also be zero, that is the set of all
Maxwell solutions, given by $A$, should
be fully covered.

The canonical map $H$ is constructed as follows. Given the equation of motion \eqref{bianchi},
locally by the Poincar\'{e} lemma, we can write $\phi$ as 

\begin{equation}
\phi \,\, = \,\, \dd \overline{a}
\label{solutiona}
\end{equation}

\noindent
with some possible ambiguity given by the addition of a closed form.
The second equation of motion, \eqref{eomgauge}, is the statement that $a$ does not contribute
to the self-dual part, hence it must describe the anti-self-dual piece. 
It becomes natural to define

\begin{equation}
A \,\, = \,\, a \, + \, \overline{a}
\label{mapele}
\end{equation}

\noindent
since it satisfies Maxwell's equations as a consequence of self-duality:

\begin{align}
\dd \star \dd A &= \dd \star \dd \left( a + \overline{a} \right) \nonumber \\
&= \dd \star \left( \dd a + \dd \overline{a} \right) \nonumber \\
&= i\dd \left( \dd a - \dd \overline{a} \right) \nonumber \\
&= 0.
\end{align}

\noindent
Notice that the kernel of \eqref{mapele} indeed vanishes. 
One takes $- a + \dd \alpha = \overline{a}$, for some $\alpha$, and, by consequence of \eqref{eomgauge}, $\phi = 0$, which 
forces $a$ to be pure gauge.
That the cokernel vanishes is a more subtle point. Because the Hodge star operator $\star$ satisfies $\star^{2} = - 1$ in
four dimensions, it splits the bundle $\Lambda^{2}$, of two-forms in Minkowski space, into a direct sum,

\begin{equation}
\Lambda^{2} \, = \, \Lambda^{2}_{+} \oplus \Lambda^{2}_{-},
\end{equation}

\noindent
where $\Lambda^{2}_{\pm}$ are the $\pm i$ eigenspaces of $\star$.
Thus, any two form can be written as

\begin{equation}
F = \phi \, + \, \overline{\phi}
\end{equation}

\noindent
and, by the Poincar\'{e} lemma, we locally have the decomposition 
\eqref{mapele}.

The analysis of this construction is special to the 4-dimensional Minkowski space
and it carries through only for the equations of motion. 
It is not true that the action \eqref{paction} is off-shell equivalent to the Maxwell action.
One way to understand this is to notice that the action \eqref{paction} is not real.
In general, equation \eqref{paction} is defined over some complex infinite-dimensional manifold.

Such consideration raises the question if whether the map \eqref{mapele} defines a real $A$ or not.
It turns out that, in phase space, complex conjugation acts as an involution,
where the complex conjugation map, denoted $\texttt{c.c.}$, is

\begin{center}
\begin{equation}
\texttt{c.c.} 
\begin{pmatrix}
a \\
\phi
\end{pmatrix}
=
\begin{pmatrix}
\dd^{-1} \overline{\phi} \\
\dd \overline{a}
\end{pmatrix}.
\label{involution}
\end{equation}
\end{center}

\noindent
It has fixed point given by

\begin{equation}
\overline{\phi_{AB}} = \overline{\phi}_{\da \db}  = \partial_{C (\da} a^{C}_{\,\,\,\,\,\,\db)},
\label{fixedpoint}
\end{equation}

\noindent
from where we see that the complex conjugate of $a$ is $\overline{a}$ and vice-versa.
To summarize our results: the action \eqref{paction} is complex, but in phase space --
that is, the space of classical solutions -- there is a well-defined notion of reality, which
is given by the fixed point of the involution \eqref{involution}, namely equation \eqref{fixedpoint}.
Only in this submanifold, the two theories classically agree.

Outside the fixed point, the complex theory describes two photons.
Self-duality of $\phi$ allows one to write

\begin{equation}
\phi = F + i \star F
\end{equation}

\noindent
for a real $2$-form $F$. Hence, the equation of motion $\dd \phi = 0$ implies
Maxwell's equations:

\begin{equation}
\dd F = 0 \quad \text{ and } \quad  \dd \star F = 0.
\end{equation}

\noindent
On the other hand, the gauge field $a$ on-shell gives an anti-self-dual $2$-form:

\begin{equation}
\dd a = G - i \star G
\end{equation}

\noindent
from where the second Maxwell equations come:

\begin{equation}
\dd G = 0 \quad \text{ and } \quad \dd \star G = 0.
\end{equation}

\noindent
The reality conditions \eqref{fixedpoint} impose
$F = G$.
\subsection{Making action real.}

Consider the real part of the action\footnote{We would like to thank Arkady Tseytlin for suggesting this idea.} \eqref{paction}:

\begin{equation}
S =  \int \left( \phi \wedge \dd a + \overline{\phi} \wedge \dd \overline{a} \right).
\label{realaction}
\end{equation}

\noindent
It turns out that the equations of motion are unchanged. To see this, 
consider the variation of this action under the real and imaginary parts of $a$, it gives

\begin{equation}
\dd \left( \phi + \overline{\phi} \right) = 0 \quad \text{ and } \quad \dd \left( \phi - \overline{\phi} \right) = 0
\end{equation}

\noindent
respectively. Self-duality of $\phi$ does not allow us to vary its real and imaginary parts 
independently, therefore we have a single equation of motion:

\begin{equation}
\dd \left( a + \overline{a} \right) + i \star \dd \left( a - \overline{a} \right) = 0.
\end{equation}

\noindent
Inspection shows that the real and imaginary parts of $a$ satisfy the Maxwell's equations while
$\phi$ again satisfies $\dd \phi = 0$.
The two copies of the Maxwell theory can be identified with 
the reality condition \eqref{fixedpoint}. It is surprising
that the addition of complex conjugatation does not change the field content of the theory.

\subsection{Symplectic structure.}

We wish to establish the above correspondence for every spin $s$ field.
The above consideration can be rephrased using the notion of symplectic structure\footnote{For a brief review of these terms, see appendix \ref{appendix}.}.
In this language, although the action is defined for complex field configurations, there is a real submanifold inside the phase space where the restriction of the symplectic form derived from \eqref{paction} 
is non-degenerate.
Then, we will construct a map $H$ that becomes a canonical transformation to the phase space of Fronsdal. 

The symplectic structure for action \eqref{paction} is 

\begin{equation}
\Omega = i \int_{C} \delta \phi^{AB \cdots \,D} \wedge \delta a^{\da}_{\,\,\,\,\,\,B \cdots \, D} \wedge \dd^{3} x_{A \da}, \quad \text{where} \quad  n_{A \da} \dd^{3} x = \dd^{3} x_{A \da},
\label{psymplectic}
\end{equation} 

\noindent
for a normal vector $n_{A\da}$ to the spacelike contour $C$.
It is $\delta$-closed and invariant under deformations of $C$, because

\begin{equation}
\partial_{A \da} \left( \delta \phi^{AB \cdots \,D} \wedge \delta a^{\da}_{\,\,\,\,\,\,B \cdots \, D} \right) = 0
\end{equation}

\noindent
once we use the equations of motion. 
However, note that this symplectic structure is also degenerate.
Degeneracies indicate the presence of gauge symmetries in
the action. In our case, if we let

\begin{equation}
V = \partial_{\da (B} \xi_{ \cdots \, D)} (x) \frac{\delta}{\delta a_{\da B \cdots D}(x)} 
\end{equation}

\noindent
be a tangent vector field along gauge trajectories, we get

\begin{align}
\iota_{V} \Omega &= i \int_{C} \partial^{\da}_{\,\,\,\,\,\,(B} \xi_{ \cdots \, D)}  \delta \phi^{AB \cdots \,D}  \dd^{3} x_{A \da} \nonumber \\
&= i \int_{C} \partial^{\da}_{\,\,\,\,\,\,(B} \left[ \xi_{ \cdots \, D)}  \delta \phi^{AB \cdots \,D} \right]  \dd^{3} x_{A \da} 
- i \int_{C}  \xi_{(C \cdots \, D}  \partial_{B)}^{\,\,\,\,\,\,\da} \delta \phi^{ABC \cdots \,D}  \dd^{3} x_{A \da} \nonumber \\
&= i \int_{C} \partial^{\da}_{\,\,\,\,\,\, (B} \left[ \xi_{ \cdots \, D)}  \delta \phi^{AB \cdots \,D} \right]  \dd^{3} x_{A \da} = 0,
\end{align}

\noindent
where the last line vanishes due to $C$ being a closed contour.
Degenerate symplectic structures 
descend to a reduced phase space. If we define $\ker \Omega$ to be the set of 
gauge generators, then the reduced phase space is given by the factor $M/\ker \Omega$.
On-shell gauge-invariant functions are points in this space and they coincide with physical
observables.

It still remains to be checked whether this symplectic structure is real
over the fixed point defined by the involution\footnote{See paragraph above equation \eqref{fixedpoint}}.
The fixed point can be written as 

\begin{equation}
\overline{\phi_{AB \cdots D}} = \overline{\phi}_{\da \db \cdots \dD} = \partial_{(\db}^{\,\,\,\,\,\,\,B} \cdots \partial_{\dD}^{\,\,\,\,\,\,D} \, a_{\da) B \cdots D}
\label{generalfixedpoint}
\end{equation}

\noindent
and it follows that

\begin{align}
\overline{\Omega} &=
-i \int_{C} \delta \overline{\phi}^{\da \db \cdots \, \dD} \wedge \delta \overline{a}^{A}_{\,\,\,\,\,\,\db \cdots\, \dD} \wedge \dd^{3} x_{A \da} \nonumber \\
&= -i \int_{C} \partial^{(\db}_{\,\,\,\,\,B} \, \partial^{\dc}_{\,\,\,\,\,C} \, \ldots \, \partial^{\dD}_{\,\,\,\,\,D} \, \delta a^{\da) B \cdots D} \wedge \delta \overline{a}^{A}_{\,\,\,\,\, \db \cdots \dD}
\wedge \dd^{3} x_{A \da} \nonumber \\
&= (-)^{2s+1} i \int_{C} 
\delta a^{\da B \cdots D} \wedge \partial^{\db}_{\,\,\,\,\,(B} \, \partial^{\dc}_{\,\,\,\,\,C} \, \ldots \, \partial^{\dD}_{\,\,\,\,\,D} \, \delta \overline{a}_{A) \db \cdots \dD}
\wedge \dd^{3} x_{\da}^{\,\,\,\,\,A} \nonumber \\
&= - i \int_{C} \delta a^{\da}_{\,\,\,\,\,\,B \cdots D} \wedge \delta \phi^{AB \cdots D} \wedge \dd^{3} x_{A \da} \nonumber \\
&= + \Omega,
\end{align}

\noindent
thus proving that indeed the symplectic structure is real.

Having the symplectic structure for Penrose theory,
it remains to construct the canonical map which will relate the two descriptions.
In doing so, we are ready to prove that the two phase spaces agree.

\section{Canonical map between descriptions.}

It is instructive to consider some examples before treating the general case.
We specialize our discussion to Rarita-Schwinger and linearized gravity in the next two subsections.
Each case will serve to emphasize the introduction of a new tool for the analysis. 

In the Rarita-Schwinger case, for example, we will see how the 
spliting of the gauge field into self-dual and anti-self-dual connection -- as it has already happened in electromagnetic case --
comes about in the symplectic structure.
The main objective is to demonstrate, on the real slice given by 
\eqref{generalfixedpoint}, that the canonical map indeed preserves the symplectic structure.

In linearized gravity, we show how the analysis can be made rather straightforward once we pass to momentum space. 
It will avoid dealing with integration by parts when we show that the symplectic structures agree.

\subsection{Rarita-Schwinger case.}

The Rarita-Schwinger theory is obtained when $h=1$ in Section \ref{sectionfronsdal}. 
We have the Majorana spinor 

\begin{equation}
\Psi_{m} = 
\begin{pmatrix}
\psi_{A \, m} \\
\overline{\psi}^{\da}_{m}
\end{pmatrix}
\end{equation}

\noindent
with higher-spin gauge symmetries $\delta \Psi_{m} = \partial_{m} \varepsilon $ and
gauge-invariant action 

\begin{equation}
S = \int \dd^{ 4} x \,\,\, \left( \overline{\Psi}^{m} F_{m} + \frac{1}{2} \,  \overline{\Psi}_{p} \Gamma^{p} \, \Gamma^{m} F_{m} \right).
\label{rarita}
\end{equation}

\noindent
The equations of motion read 

\begin{equation}
F_{m} = \Gamma^{n} \partial_{n} \Psi_{m} - \partial_{m} \Gamma^{n} \Psi_{n} = 0.
\label{sec3eq2}
\end{equation}

For our applications, it will be useful to 
consider the gauge-invariant combination 

\begin{equation}
R_{mn} = \partial_{m} \Psi_{n} - \partial_{n} \Psi_{m},
\label{raritacurvature}
\end{equation}

\noindent
in order to make contact with the curvature spinors $\phi^{ABC}$ and $\overline{\phi}^{\da \db \dc}$.
To see how, let us introduce the following spinor counterpart of $R_{mn}$:

\begin{equation}
R_{M \dM  N \dN} =  \dd \Psi_{(\dM \dN) }  \ep_{MN} - \dd \Psi_{(M N)} \ep_{\dM \dN},
\end{equation}

\noindent
where abbreviations have been used: 

\begin{subequations}
\begin{equation}
\partial^{A}_{\,\,\,\,\,\, (\dN} \Psi_{\dM) A} =  \dd \Psi_{(\dM \dN) }  =
\begin{pmatrix}
\dd \psi_{(\dM \dN) B} \\
\dd \overline{\psi}^{\db}_{\,\,\,\,\,\, (\dM \dN)}
\end{pmatrix}
\end{equation}

\noindent
and 

\begin{equation}
\partial_{\da (M} \Psi_{N)}^{\,\,\,\,\,\, \da} = \dd \Psi_{(M N)} = 
\begin{pmatrix}
\dd \psi_{(MN)B} \\
\dd \overline{\psi}^{\db}_{\,\,\,\,\,\,(MN)}
\end{pmatrix}.
\end{equation}
\label{sec4eq7}
\end{subequations}

\noindent
It enables us to rewrite the equations of motion in the form

\begin{equation}
\Gamma^{m}R_{mn} = 0 \longmapsto
\begin{pmatrix}
0 & \delta^{\,\,\,\,\,M}_{B} \delta^{\dM}_{\,\,\,\,\,\db} \\
\ep^{MB} \ep^{\dM \db} & 0 
\end{pmatrix}
\begin{pmatrix}
\dd \psi_{\dM \dN B} \ep_{MN} - \dd \psi_{MNB} \ep_{\dM \dN} \\
\dd \overline{\psi}^{\db}_{\,\,\,\,\,\, \dM \dN} \ep_{MN}  - \dd \overline{\psi}^{\db}_{\,\,\,\,\,\,MN} \ep_{\dM \dN}
\end{pmatrix}
=0.
\label{map}
\end{equation}

%Andrei wrote: Something seriously wrong with indices. I don't see where.

\noindent
from where we obtain 

\begin{subequations}
\begin{equation}
\dd \psi_{\da \dN N} - \dd \psi^{C}_{\,\,\, N C} \, \ep_{\da \dN} =0
\label{aqui}
\end{equation}

\noindent
and

\begin{equation}
\dd \overline{\psi}^{\dc}_{\,\,\,\,\,\, \dc \dN} \ep_{AN}  + \dd \overline{\psi}^{\dN}_{\,\,\,\,\,\,AN}  = 0.
\label{aqui2}
\end{equation}
\end{subequations}

\noindent
A quick inspection shows the only possible solutions for \eqref{aqui} are

\begin{equation}
\dd \psi_{\da \dN N} = 0 \quad \text{ and } \quad \dd \psi^{C}_{\,\,\, N C} = 0
\label{analise}
\end{equation}

\noindent
since the first term is symmetric in $\da \dN$ while the second one is anti-symmetric in $\da \dN$.
The same type of reasoning leads us to the solutions of \eqref{aqui2}:

\begin{equation}
\dd \overline{\psi}^{\dc}_{\,\,\,\,\,\, \dc \dN} = 0 \quad \text{ and } \quad \dd \overline{\psi}^{\dN}_{\,\,\,\,\,\,AN} = 0.
\end{equation}

\noindent
These solutions
annihilate any components with dotted and undotted indices. Moreover
they completely symmetrize the self-dual and anti-self-dual part.
The remaining components split $R_{mn}$ into 

\begin{equation}
R_{mn} \longmapsto - \dd \psi_{(MNA)} \ep_{\dM \dN} - \dd \overline{\psi}_{(\da \dM \dN)} \ep_{M N}
\end{equation}

\noindent 
and we can identify

\begin{equation}
- \dd \psi_{(MNA)} \quad  \text{ as } \quad  \phi_{AMN},
\end{equation}

\noindent
and 

\begin{equation}
- \dd \overline{\psi}_{\da \dM \dN} \quad \text{ as } \quad \overline{\phi}_{\da \dM \dN}.
\end{equation}

This procedure occurs for other spins as well.
One defines a gauge-invariant combination,
and once the equations of motion are imposed the spinors $\phi^{AB \cdots D}$ and
$\overline{\phi}^{\da \db \cdots \dD}$ are the only remaining components. 
Notice that 

\begin{equation}
\partial_{[m} R_{np]} = 0
\end{equation}

\noindent
is trivially satisfied in the presence of $\Psi_{m}$. As soon as we change pictures and use
the curvature spinors, this equation turns into an equation of motion. The anti-symmetry is equivalent
to a contraction of spinor indices, and so we recover \eqref{bianchi} and 
\eqref{bianchi2}:

\[
\partial^{\da A} \phi_{A MN} = 0 \quad \text{ and } \quad  \partial^{A \da} \overline{\phi}_{\da \dM \dN} = 0.
\]

\noindent
The Penrose description splits the gauge field $h_{m_{1} \cdots m_{s}}$ into
anti-self-dual and self-dual parts treating the self-dual part via the curvature while the anti-self-dual part
is described with the anti-self-dual gauge field.

In the Rarita-Schwinger case, the gauge field $a_{\da B C}$ is mapped to the anti-chiral part $\overline{\psi}^{m \da}$
with the \textit{ansatz} 

\begin{equation}
\overline{\psi}^{m \da} = i \overline{\sigma}^{m \dot{E} E} \left( \partial^{\da C} a_{\dot{E} C E} + \frac{1}{2} \partial_{\dot{E}}^{\,\,\,\,\,C} a^{\da}_{\,\,\,\, CE} \right)
\label{ansatz}
\end{equation}

\noindent
where the coefficients are fixed by requiring the higher-spin gauge symmetries to coincide. For consistency,
it is also possible, with this choice, to check that $\overline{\psi}^{m \da }$ satisfies the equations of motion
when $a_{\da BC}$ does. 
We should point out that this map is the non-trivial piece of our correspondence.
For other higher-spins, it has to be constructed with the right coefficients case by case.

%Write about when is possible to do this.

One can derive the symplectic structure from action \eqref{rarita}
and it reads:

\begin{align}
\Omega = \int \, &\left( 2 \delta \psi_{m} \wedge \sigma^{m} \overline{\sigma}^{np} \delta \overline{\psi}_{p} + 
2 \delta \overline{\psi}_{m} \wedge \overline{\sigma}^{m} \sigma^{np} \delta \psi_{p} \right. \nonumber \\
&+ \left.\delta \psi_{m} \wedge \sigma^{n} \delta \overline{\psi}^{m} + \delta \overline{\psi}_{m} \wedge \overline{\sigma}^{n} \delta \psi^{m} - \delta \psi^{n} \wedge \sigma^{m} \delta \overline{\psi}_{m} 
- \delta \overline{\psi}^{n} \wedge \overline{\sigma}^{m} \delta \psi_{m}\right) \wedge d^{\, 3} x_{n}.
\label{sec3eq51}
\end{align}

\noindent
If we intend to describe the spin $3/2$ piece, we are allowed to use the gauge

\begin{equation}
\Gamma^{m} \Psi_{m} = 0
\label{lorentz}
\end{equation}

\noindent
so the symplectic structure collapses to

\begin{equation}
\Omega = 2 \int \, \delta \psi_{m} \wedge \sigma^{n} \delta \overline{\psi}^{m} \wedge \dd^{\,3} x_{n}.
\label{symplecticraritaf}
\end{equation}

\noindent
In Penrose case, the symplectic structure follows from
\eqref{paction}, and it is

\begin{equation}
\Omega = i \int \delta \phi^{ABC} \wedge \delta a^{\da}_{\,\,\,\,\, BC} \wedge \dd^{\, 3} x_{A \da}.
\label{symplecticraritap}
\end{equation}

\noindent
Notice the gauge condition \label{lorentz} implies 

\begin{equation}
\partial^{\db A} \psi_{\dot{E} E A} =0,
\label{usei}
\end{equation}

\noindent
and by consequence of \eqref{ansatz}:

\begin{equation}
\partial_{\da}^{\,\,\,\,\,B} \, a^{\da}_{\,\,\,\,\,\, B C} = 0.
\label{graviansatz}
\end{equation}

When substitute our \textit{ansatz} into the symplectic structure \eqref{symplecticraritaf},
we obtain

\begin{equation}
\Omega = + i \int \delta \psi^{\dot{E} E A} \wedge  \partial_{\dot{E}}^{\,\,\,\,\,\,C}  \delta a_{\da EC}  \wedge \dd^{3} x_{A}^{\,\,\,\,\,\,\da}
\label{meiodocaminho}
\end{equation}

\noindent
and there is a subtlety we must highlight. 
Despite the advantage of being able to use the equations of motion when dealing with a symplectic
structure, we are not allowed to integrate by parts indiscriminately.
If we assume, for the moment, that we can make such integration, then we would get
the desired result:

\begin{equation}
\Omega = + i \int \delta \psi^{\dot{E} E A} \wedge  \partial_{\dot{E}}^{\,\,\,\,\,\,C}  \delta a_{\da EC}  \wedge \dd^{3} x_{A}^{\,\,\,\,\,\,\da}
= -i \int \partial_{\dot{E}}^{\,\,\,\,\,\,C} \delta \psi^{\dot{E} E A} \wedge \delta a_{\da EC}  \wedge \dd^{3} x_{A}^{\,\,\,\,\,\,\da},
\label{sec6eq8}
\end{equation}

\noindent
because, by the equations of motion, the $\dd \psi$ term is symmetric in the pair $CE$ but also in $EA$
-- thus being symmetric in all of its 
indices -- and we have

\begin{equation}
\Omega = - i \int \partial_{\dot{E}}^{\,\,\,\,\,\,C} \delta \psi^{\dot{E} E A} \wedge \delta a_{\da EC}  \wedge \dd^{3} x_{A}^{\,\,\,\,\,\,\da}
= -i \int \, \delta \phi^{CEA} \wedge \delta \, a_{\da EC} \wedge \dd^{3} x_{A}^{\,\,\,\,\,\, \da}.
\label{sec6eq9}
\end{equation}

The integration by parts is justified if we show that the two terms differ by an exact form.
Consider

\begin{align}
& \int \partial_{m} \, \delta X^{[mn]} \wedge \dd^{3} x_{n} = 
\int \partial_{\dot{E} C} \, \delta X^{ [ \dot{E} C | \da A ]} \wedge \dd^{3} x_{\da A} \nonumber \\
&= - \int \partial_{\dot{E}}^{\,\,\,\,\,\,C}  \left( \delta \psi^{\dot{E} A E} \wedge \delta a_{\da CE} 
- \delta \psi^{\,\,\,\,\,\, E}_{\da\,\,\,\,\,\,\,C} \wedge \delta a^{\dot{E} A}_{\,\,\,\,\,\,\,\,\,\,E} \right) 
\wedge \dd^{3} x_{A}^{\,\,\,\,\,\,\da}
\label{parts}
\end{align}

\noindent
and notice that \eqref{parts} is exactly what we want:

\begin{equation}
- \int \left (\partial_{\dot{E}}^{\,\,\,\,\,\,C} \delta \psi^{\dot{E} E A} \wedge \delta a_{\da EC}  \wedge \dd^{3} x_{A}^{\,\,\,\,\,\,\da}
+ \delta \psi^{\dot{E} E A} \wedge  \partial_{\dot{E}}^{\,\,\,\,\,\,C}  \delta a_{\da EC}  \wedge \dd^{3} x_{A}^{\,\,\,\,\,\,\da} \right),
\label{sec7eq3}
\end{equation}

\noindent
since all other terms cancel after we use \eqref{usei} together with the equation of motion for the gauge field $a_{\da B \cdots D}$:

\begin{equation}
\partial^{\da}_{\,\,\,\,\,( A} a_{BC) \da} = 0.
\end{equation} 

\noindent
In all other cases, the integration by parts will be the main issue.
We circumvent the difficulty of finding appropriate exact forms 
by working in momentum space.

\subsection{Linearized gravity case.}

When $s=2$ in section \ref{sectionfronsdal} we have linearized
Einstein theory of gravity. The field $h_{mn}$ has 
gauge invariance of the form

\begin{equation}
\delta_{\xi} h_{mn} (x) = \partial_{m} \xi_{n} (x)  + \partial_{n} \xi_{m} (x) 
\label{hgauge}
\end{equation}

\noindent
and is described by the flat space action

\begin{equation}
S = -\frac{1}{2} \int \dd^{4} x \, \left( h^{mn} R_{mn} - \frac{1}{2} h^{p}_{\,\,\,\,p} \, R^{q}_{\,\,\,\,q} \right).
\end{equation}

\noindent
The $R_{mn}$ and $R^{p}_{\,\,\,\,p}$ represent the Ricci tensor and Ricci scalar respectively.
Both can be obtained from the linearized curvature given by

\begin{equation}
R_{mnpq} = 4 \, \partial_{[m} h_{n][p} \overleftarrow{\partial}_{q]}.
\end{equation}

\noindent
The equations of motion are the linearized Einstein field equations

\begin{equation}
R_{mn}= 0
\end{equation}

\noindent
and the symplectic structure is 

\begin{align}
\Omega = - \frac{1}{2}
\int & \left( 2 \delta h^{m}_{\,\,\,\,n} \wedge \partial^{p} \delta h_{p}^{\,\,\,\,n} - \delta h_{pn} \wedge \partial^{m} \delta h^{pn} + \delta h_{p}^{\,\,\,\,p} \wedge \partial^{m} \delta h_{n}^{\,\,\,\,n}
\right. \nonumber \\
&- \left. \partial_{n} \delta h^{mn} \wedge \delta h_{p}^{\,\,\,\,p} + \partial^{p} \delta h^{n}_{\,\,\,\,n} \wedge \delta h_{p}^{\,\,\,\,m} \right) \wedge \dd^{3} x_{m}.
\label{sec9eq3}
\end{align}

In order to change to Penrose description, we 
need to identify the $(\phi,a)$ fields.
The self-dual part of $R_{mnpq}$ gives $\phi_{MNPQ}$ via

\begin{equation}
\phi_{MNPQ} = \partial_{\dM(M} \partial_{|\dN| N} h^{\dM \dN}_{\,\,\,\,\,\,\,\,\,\,\,\, P Q)},
\end{equation}

\noindent
while the anti-self-dual piece is described by the map

\begin{equation}
h_{ M \dM  N \dN } =  -i\partial_{\dM}^{\,\,\,\,\,\,\,C} a_{\dN C M N}   -i\partial_{\dN}^{\,\,\,\,\,\,C} a_{\dM C M N}.
\label{map}
\end{equation}

\noindent
Again, \eqref{map} is an \textit{ansatz}. It is constructed by requiring gauge symmetries to coincide.
An interesting feature we should stress is that $h$ comes traceless since $a$ is completely symmetric in
its undotted indices. This is not a problem. In Fronsdal theory
these degrees of freedom are pure gauge. 

%But it is intructive to see how the gauge parameters are related.
%Given \eqref{hgauge} and
%\begin{equation}
%\delta_{\zeta} a_{\dM N P Q} = \partial_{\dM ( N} \zeta_{PQ)}
%\end{equation}
%\noindent
%we obtain
%\begin{equation}
%\xi_{M \dM} = 2 \partial_{\dM}^{\,\,\,\,C} \zeta_{C M}
%\end{equation}

We will demonstrate that the phase spaces of these descriptions agree. 
In this on-shell counting, let us go into Fourier space and fix the only non-zero component
of the momentum to be $p_{2}^{\,\,\,\, \dot{2}}$.
From the spinor description, we have then

\begin{equation}
\partial^{\da A} \phi_{ABCD} = 0 \quad \Longrightarrow \quad p^{\dot{2} 1} \phi_{1 BCD} = 0,
\label{sec10eq1}
\end{equation}

\noindent
which implies that every term with an $1$ index vanishes. 
The only non-zero component of $\phi$ thus is $\phi_{2222}$.
For the gauge field $a$, we have

\begin{equation}
\partial_{(A}^{\,\,\,\,\,\,\da} a_{ BCD) \da} = 0 \quad \Longrightarrow \quad p_{(2}^{\,\,\,\,\,\, \dot{2}} a_{ BCD) \dot{2}} = 0,
\label{sec10eq2}
\end{equation}

\noindent
which means that every $a$ with a $\dot{2}$ and a $2$ index vanishes. 
The only remaining degrees of freedom are $a_{\dot{1} BCD}$.
However, we should account for the gauge invariance:

\begin{equation}
\delta a_{\da BCD} = \partial_{\da (B} \, \xi_{CD)} \quad \Longrightarrow \quad \delta a_{\dot{1} 2 CD} = p_{\dot{1} (2} \, \xi_{CD)},
\label{sec10eq3}
\end{equation}

\noindent
which makes the only non-zero component  $a_{\dot{1} 111}$. Finally the symplectic structure 
for spin 2 Penrose theory is

\begin{equation}
\Omega = i \int \delta \phi^{1111} \wedge \delta a^{\dot{2}}_{\,\,\,\,\,\, 111} \wedge \dd^{3} x_{1 \dot{2}}.
\label{sec10eq4}
\end{equation}

Let us turn to Fronsdal theory.
Fix a gauge where $h_{mn}$ is traceless, 
so the symplectic structure \eqref{sec9eq3} reduces to

\begin{equation}
\Omega = - \frac{1}{2}
\int \left( 2 \delta h^{m}_{\,\,\,\,n} \wedge \partial^{p} \delta h_{p}^{\,\,\,\,n} - \delta h_{pn} \wedge \partial^{m} \delta h^{pn} \right) \wedge \dd^{3} x_{m}.
\label{sec10eq5}
\end{equation}

\noindent
The degrees of freedom of the self-dual part are fixed by Einstein's equation
since $\phi$ is written in terms of $h$. For spin 2:

\begin{equation}
R_{(M \dM | N \dN)} = p^{2} h_{(M \dM | N \dN)} + p_{ ( M \dM} p^{a} h_{| a | \, N \dN)} = 0,
\label{sec10eq7}
\end{equation} 

\noindent
which gives, after we impose $p^{2}=0$,

\begin{equation}
p_{ (M \dM} h_{ N \dN) 1 \dot{2}} = 0.
\label{sec10eq8}
\end{equation}

\noindent
The general solution of this equation is 

\begin{equation}
h_{( 1 \dot{2} | M \dM ) } = 0.
\label{sec10eq9}
\end{equation}

\noindent
So, for the self-dual part of the curvature, we have then

\begin{equation}
\phi_{22 CD} = - p_{(2}^{\,\,\,\,\,\, \dot{2}} p_{2}^{\,\,\,\,\,\, \dot{2}} h_{CD) \dot{2} \dot{2}} \quad \Longrightarrow \quad
\phi_{2222} = - p_{2}^{\,\,\,\,\,\, \dot{2}} p_{2}^{\,\,\,\,\,\, \dot{2}} h_{22 \dot{2} \dot{2}}.
\label{sec10eq10}
\end{equation}

To connect the two descriptions, we split the gravitational field $h$ into a self-dual and anti-self-dual part.
The self-dual piece is already described by Einstein's equations while the anti-self-dual part is given by
the \textit{ansatz} \eqref{graviansatz}. It implies:

\begin{equation}
h_{( 1 \dot{1} | 1 \dot{1} )} =  p_{\dot{1}}^{\,\,\,\,1} a_{\dot{1} 1 1 1} +  p_{\dot{1}}^{\,\,\,\,1} a_{\dot{1} 1 1 1} = + 2 p_{\dot{1}}^{\,\,\,\,1} a_{\dot{1} 1 1 1}.
\label{sec10eq6}
\end{equation}

\noindent
These considerations collapse the symplectic structure to

\begin{align}
\Omega &= - \frac{i}{2} \int \left( 2 \delta h^{m}_{\,\,\,\,n} \wedge p^{1\dot{2}} \delta h_{1 \dot{2}}^{\,\,\,\,\,\,\,\,n} \right) \wedge \dd^{3} x_{m}
- \left( \delta h_{pn} \wedge p^{1 \dot{2}} \delta h^{p n} \right) \wedge \dd^{3} x_{1 \dot{2}} \nonumber \\
&= + \frac{i}{2} \int \left( \delta h_{pn} \wedge p^{1 \dot{2}} \delta h^{p n} \right) \wedge \dd^{3} x_{1 \dot{2}} \nonumber \\
&= + i \int \left( p_{\dot{1}}^{\,\,\,\,\,\, 1} \delta a_{\dot{1} 111}  \wedge p^{1 \dot{2}} \delta h^{\dot{1} 1\dot{1}1} \right) \wedge \dd^{3} x_{1 \dot{2}} \nonumber \\
&= + i \int \left(  \delta a_{\dot{1} 111}  \wedge p_{\dot{1}}^{\,\,\,\,\,\, 1} p^{1 \dot{2}} \delta h^{\dot{1} 1\dot{1}1} \right) \wedge \dd^{3} x_{1 \dot{2}} \nonumber \\
&= - i \int \left(  \delta a_{\dot{1} 111}  \wedge p_{2}^{\,\,\,\,\,\, \dot{2}} p_{2}^{\,\,\,\,\,\, \dot{2}} \delta h_{22 \dot{2} \dot{2}} \right) \wedge \dd^{3} x_{1 \dot{2}} \nonumber \\
&= + i  \int \left( \delta a_{\dot{1} 111}  \wedge \delta \phi_{2222}  \right) \wedge \dd^{3} x_{1 \dot{2}}.
\label{sec10eq11}
\end{align}

This computation highlights the usefulness of momentum space.
We can work directly with physical degrees of freedom as it is suggested 
when dealing with symplectic structures.

\subsection{Canonical map between formulations for general spin s.}

In order to relate the two descriptions in general case, we split the Fronsdal field $h_{m_{1} \cdots \, m_{s}}$ into self-dual and anti-self-dual components.
The anti-self-dual part is described by the gauge field $a_{\dM A \cdots \, N}$ via

\begin{equation}
h_{M_{1} \dM_{1}  \, \cdots \,  M_{s} \dM_{s}} =  (-i)^{2s-1} \partial_{ ( \dM_{s}}^{\,\,\,\,\,\,\,\, N_{s}} \ldots \, \partial_{\dM_{2}}^{\,\,\,\,\,\,\, N_{2}} 
\, a_{\dM_{1} ) N_{2} \cdots N_{s} M_{1} \cdots M_{s} },
\label{amaptoh}
\end{equation}

\noindent
while the self-dual degrees of freedom are given by the curvature $\phi_{A \cdots \, D}$, which should come from the gauge-invariant tensor

\begin{equation}
R_{[m_{1} n_{1}] \cdots \, [m_{s} n_{s}]} = \partial_{[n_{s} |} \partial_{[n_{s-1} |} \ldots \, \partial_{ |[ n_{1}} h_{m_{1}] | \cdots \, | m_{s-1}] | m_{s}]}.
\label{sec11eq2}
\end{equation}

\noindent
Once Fronsdal equations are imposed, we expect\footnote{Remember, to a spacetime index $m$ there corresponds a pair $M \dM$.}

\begin{equation}
\phi_{M_{1}N_{1} \cdots M_{s} N_{s}} = R_{ (M_{1}N_{1} \, \cdots \, M_{s} N_{s})} = \partial_{ (N_{s}}^{\,\,\,\,\,\,\, \dN_{s}} \ldots \, \partial_{N_{1}}^{\,\,\,\,\,\,\,\, \dN_{1}} 
\, h_{M_{1} \ldots \, M_{s})  \dN_{1} \ldots \, \dN_{s}}.
\end{equation}

\noindent
We also expect that any component of $R_{m_{1} n_{1} \cdots \, m_{s} n_{s}}$ which contains mixed dotted and undotted indices should vanish.
In what follows, we will prove that this is indeed the case.

%Its spinor counterpart is schematically written as
%\begin{align}
%R_{[m_{1} n_{1}] \cdots \, [m_{s} n_{s}]} &= R_{(M_{1} N_{1}) \, \cdots \, (M_{s} N_{s})} \nonumber \\
%&+ R_{(\dM_{1} \dN_{1}) (M_{2} N_{2}) \, \cdots \, (M_{s} N_{s})} + R_{(M_{1} N_{1}) (\dM_{2} \dN_{2}) \, \cdots \, (M_{s} N_{s})} + \,\ldots \,+ R_{(M_{1} N_{1}) (M_{2} N_{2}) \, \cdots \, (\dM_{s} \dN_{s})}
%\nonumber \\
%&+ R_{(\dM_{1} \dN_{1}) (\dM_{2} \dN_{2}) \, \cdots \, (M_{s} N_{s})} + \, \ldots \, + R_{(M_{1} N_{1})  \, \cdots \, ( \dM_{s-1} \dN_{s-1})( \dM_{s} \dN_{s})} \nonumber \\
%& \, \, \,. \nonumber \\
%& \, \, \,. \nonumber \\
%& \, \, \,. \nonumber \\
%&+  R_{(\dM_{1} \dN_{1}) \, \cdots \, (\dM_{s} \dN_{s})}
%\label{sec11eq3}
%\end{align}
%\noindent
%where, for example,
%\begin{equation}
%R_{ (M_{1}N_{1}) \, \cdots \, (M_{s} N_{s})} = \partial_{ N_{s}}^{\,\,\,\,\,\,\, \dN_{s}} \ldots \, \partial_{N_{1}}^{\,\,\,\,\,\,\,\, \dN_{1}} 
%\, h_{M_{1} \ldots \, M_{s}  \dN_{1} \ldots \, \dN_{s}}.
%\label{sec11eq4}
%\end{equation}

For the moment, we should stress interesting features of this map. The anti-self-dual component gives a traceless $h_{m_{1} \cdots m_{s}}$.
But this is not a problem since these degrees of freedom are pure gauge.
Moreover, in order to show that the symplectic structures match, one does not need all coefficients in the anti-self-dual map.
The Fronsdal equations will restrict these to a single component each. 
%However,
%as in the Rarita-Schwinger case,
%the right coefficients can be obtained by comparing gauge transformations.

%A probable consequence is that we will not be able to map actions, because in some sense this map relates only phase spaces.
%MELHORAR ISSO!

%It remains to show that Fronsdal equations of motion constrain these degrees of freedom leaving only the curvature spinors $\phi_{AB \cdots D}$ and $\overline{\phi}_{\da \db \cdots \dD}$.
%It turns out the cancelations happen exactly the same way as in spin $2$ case, and so we revisit it.

\subsection{Equivalent symplectic structures: Fourier counting.}

We proceed to the symplectic structures.
We circumvent the need to look for exact forms by going to momentum space, which
also makes straightforward to work only with physical degrees of freedom.

Let us choose a non-zero $p_{2}^{\,\,\,\,\,\, \dot{2}}$ component. 
Hence, the equation of motion for $a_{\da B \cdots D}$ collapses into

\begin{equation}
p_{(2}^{\,\,\,\,\,\, \dot{2}} a_{\dot{2} B_{2} \cdots \, B_{2s})}  = 0,
\label{sec111eq1}
\end{equation}

\noindent
and we can see the only non zero component is $a_{\dot{1} B \cdots D}$.
We can restrict further using the gauge transformations:

\begin{equation}
\delta a_{\dot{1} 2 \cdots D} = p_{\dot{1} (2} \xi_{ \cdots D)},
\end{equation}

\noindent
from where the only physical component which remains is $a_{\dot{1} 1 \cdots 1}$.
Thus, the map we described in \eqref{amaptoh} gives
$h_{ 1\dot{1}  \, \cdots \,  1 \dot{1} }$ component of the Fronsdal gauge field.

The degrees of freedom which the curvature spinor describes are obtained from the Fronsdal equation.
Together with the condition $p^{2} = 0$, they imply

\begin{equation}
p_{(M_{1} \dM_{1}} h_{ 1 \dot{2} M_{3} \dM_{3} \cdots \, M_{s}\dM_{s})} = 0,
\label{sec111eq2}
\end{equation}

\noindent
since our map describes a traceless $h_{m_{1} \cdots m_{s}}$ field.
This equation forces $h_{1 \dot{2} ....} = 0$,
which also annihilates any component with mixed dotted and undotted indices, and so we have

\begin{equation}
\phi_{ 22  \cdots 22 } = i^{s} p_{2}^{\,\,\,\,\, \dot{2}} \ldots \, p_{2}^{\,\,\,\,\, \dot{2}} 
\, h_{2 \ldots  2  \dot{2} \ldots \dot{2}}.
\label{sec111eq3}
\end{equation}

\noindent
Such considerations are in line with the usual formulation of Fronsdal theory,
where the degrees of freedom contained in the trace and divergence of $h_{m_{1} \cdots m_{s}}$ can be gauged away.
%Hence it is suggestive to say that, although covariant, Fronsdal formulation contains redudant degrees of freedom.

We combine all of such considerations to show the symplectic structures agree.
Note that we are allowed to discard terms of the type

\[
\int  \delta h_{....} \wedge \partial^{p} h_{p ...} \quad \text{ and } \quad \int \delta h^{p}_{\,\,\,\,p ....} \wedge \delta h_{.....}
\]

\noindent
because $h_{1 \dot{2} .....}$ vanishes and our canonical map gives a traceless $h_{m_{1} \cdots m_{s}}$.
Thus the only allowed combination for the bosonic case is of the form

\begin{equation}
\Omega = \int \left( \delta h_{n_{1} \cdots \, n_{s}} \wedge \partial^{m} \delta h^{n_{1} \cdots \, n_{s}} \right) \wedge \dd^{3} x_{m}
\label{sec111eq4}
\end{equation}

\noindent
and if we apply our results to \eqref{sec111eq4} we obtain

\begin{align}
\Omega &= \int \left( \delta h_{(1 \dot{1} | \cdots | 1 \dot{1})} \wedge p_{2}^{\,\,\,\,\, \dot{2}} \delta h^{(1 \dot{1} | \cdots | 1 \dot{1})} \right) \wedge \dd^{3} x_{\dot{2}}^{\,\,\,\,\,2}
\nonumber \\
&= (-i)^{s-1} \int \left( p^{1}_{\,\,\,\,\,\dot{1}} \ldots p^{1}_{\,\,\,\,\,\dot{1}} \delta a_{\dot{1} 1 \cdots 1} \wedge p_{2}^{\,\,\,\,\, \dot{2}} \delta  h_{2 \cdots 2  \dot{2} \cdots \dot{2}} \right)
\wedge \dd^{3} x_{\dot{2}}^{\,\,\,\,\,2} \nonumber \\
&= (-)^{s-1} (-i)^{s-1} \int \left( \delta a_{\dot{1} 1 \cdots 1} \wedge p_{2}^{\,\,\,\,\, \dot{2}} \ldots p_{2}^{\,\,\,\,\, \dot{2}} \delta  h_{2 \cdots 2  \dot{2} \cdots \dot{2}} \right)
\dd^{3} x_{\dot{2}}^{\,\,\,\,\,2} \nonumber \\
&= (-)^{s-1} (-i)^{s-1} (-i)^{s} \int \left( \delta a_{\dot{1} 1 \cdots 1} \wedge \delta \phi_{ 22  \cdots 22 } \right) \wedge \dd^{3} x_{\dot{2}}^{\,\,\,\,\,2} \nonumber \\
&= -i \int \left( \delta a_{\dot{1} 1 \cdots 1} \wedge \delta \phi_{ 22  \cdots 22 } \right) \wedge \dd^{3} x_{\dot{2}}^{\,\,\,\,\,2}
\label{sec111eq5}
\end{align}

\noindent
thus proving the desired result.
%RESULT FOR FERMIONS.
%The fermionic case is obtained with minor changes. Given the Fronsdal tensor $\Psi_{m_{1} \cdots m_{h}}$, which
%describes a spin $s=h + 1/2$ particle, we map the anti-chiral part to the gauge field:

%\begin{equation}
%\overline{\psi}_{\da \, M_{1} \dM_{1}  \, \cdots \,  M_{h} \dM_{h}} =  \partial_{  \dM_{h}}^{\,\,\,\,\,\,\,\, N_{h}} \ldots \, \partial_{\dM_{1}}^{\,\,\,\,\,\,\, N_{1}} 
%\, a_{\da N_{1} \cdots N_{s} M_{1} \cdots M_{s} },
%\end{equation}

%\noindent
%while the chiral-part describes the curvature spinor

%\begin{equation}
%\phi_{M_{1}N_{1} \cdots M_{h}N_{h}}
%\end{equation}

\section{Conformal Invariance.}

The conformal generator $\vv^{c}$ is 

\begin{equation}
\mathrm{v}^{c} = a^{c} + \omega^{cb} x_{b} + \alpha x^{c} + 2 \left( \rho \, . \, x \right) x^{c} - \rho^{c} ( x \, . \, x ),
\label{conformalgenerator}
\end{equation}

\noindent
where the first two terms are the usual Poincaré transformations; the third one describes dilatations and the last two generate
special conformal transformations.

%\subsection{Conformal invariance in linearized gravity.}
%Fronsdal theory fails to be conformal invariant at spin $2$,
%even the equations of motion. If
%we consider special conformal transformations, for example, the non-vanishing terms 
%are
%\begin{equation}
%\delta_{\vv} R_{ab} = 4 \rho_{(b} \partial_{a)} h_{c}^{\,\,\,\,c} - 4 \rho_{(b} \partial^{c} h_{a) c}
%+4 \rho^{c} \partial_{(a} h_{b) c} + 2 \eta_{ab} \biggl( \rho^{c} \partial_{c} h_{d}^{\,\,\,\,d} - 2 \rho^{c} \partial_{d} h^{d}_{\,\,\,\,c} \biggr)
%\end{equation}

\subsection{Lie derivation of spinors.}

In treating Penrose action, we are going to need to vary spinor fields under conformal transformations.
The Lie derivative of a spinor field is not widely used when compared with the usual tensor variations.
This subsection explains briefly this terminology before applying it to our case.

In geometry, given a vector field $\vv^{c}$
and a vector density $u^{b}$, the Lie derivative of $u^{b}$ with respect to $\vv^{c}$ is defined as

\begin{equation}
\mathcal{L}_{\vv} u^{b} = \vv^{a} \partial_{a} u^{b} - u^{a} \partial_{a} \vv^{b} + \mathrm{w}_{u} \, \left( \partial_{a} \vv^{a} \right) \, u^{b},
\label{usuallie}
\end{equation}

\noindent
where $\mathrm{w}_{u}$ is the density weight of $u^{b}$.
When $u^{b}$ is null, it can be written as product of two spinors, $u^{b} = \mu^{B} \overline{\mu}^{\db}$,
and so we can use equation \eqref{usuallie} 
to define the Lie derivative of $\mu^{B}$.

Following this procedure, a general spinor density\cite{cecile, tod} $\mu^{A}$ flows along the flux of $\vv^{c}$ such that its infinitesimal change is given by

\begin{equation}
\delta_{\vv} \mu^{A} = \mathcal{L}_{\vv} \mu^{A} = \vv^{m} \partial_{m} \mu^{A} + \mu^{B} f^{A}_{\,\,\,\,\,\,B} + \mathrm{w}_{\mu} \left( \partial_{m} \vv^{m} \right) \mu^{A};
\label{liespinor}
\end{equation}

\noindent
in here $\mathrm{w}_{\mu}$ denotes the density weight of the $\mu$ field and $f^{A}_{\,\,\,\,\,\,B}$ is the self-dual part of $\vv^{c}$:

\begin{equation}
f_{AB} = - \frac{1}{2} \partial_{\dc (A} \vv_{B)}^{\,\,\,\,\,\,\dc}.
\end{equation}

\noindent
In deriving \eqref{liespinor} from \eqref{usuallie}, we must impose that $\vv^{c}$ is a conformal generator.
Indeed, the second term in \eqref{usuallie} gives a contribution of the form:

\begin{align}
- u^{a} \partial_{a} \vv_{b} &= - \mu^{A} \overline{\mu}^{\da} \partial_{A \da} \vv_{B\db} \nonumber \\
&= - \mu^{A} \overline{\mu}^{\da} \partial_{[ A \da} \vv_{B\db ] } - \mu^{A} \overline{\mu}^{\da} \partial_{( A \da} \vv_{B \db)} \nonumber \\
&= - \mu^{A} \overline{\mu}^{\da} \left( f_{AB} \ep_{\da \db} + \overline{f}_{\da \db} \ep_{AB} \right) - \mu^{A} \overline{\mu}^{\da} \partial_{( A \da} \vv_{B \db)} \nonumber \\
&=  \overline{\mu}_{\db} \, \mu^{A} f_{AB}  +  \mu_{B} \, \overline{\mu}^{\da} \overline{f}_{\da \db} - \mu^{A} \overline{\mu}^{\da} \partial_{( A \da} \vv_{B \db)},
\end{align}

\noindent
in which the last term does not split into something dependent of $B$ and $\db$ separately.
It is precisely when $\vv^{c}$ is a conformal generator, that is

\begin{equation}
\partial_{(A \da} \, \vv_{B \db)} \,\, =\,\, \left(\frac{1}{2} \partial_{m} \vv^{m} \right) \ep_{AB} \, \ep_{\da \db}.
\end{equation}

\noindent
that we can identify the desired contributions to each spinor.

In our applications, of special interest is the self-dual part of the special conformal transformations.
We write it explicitly for future use:

\begin{equation}
f_{AB} = - 2 \, \rho_{\dc (A} x_{B)}^{\,\,\,\,\,\, \dc} .
\label{mistake}
\end{equation}

\subsection{Weight conventions.}

The weight of a density is a geometrical quantity, that is, it has fixed value independent of which 
transformation is made; and usually we would have

\begin{equation}
\mathcal{L}_{\vv} \ep_{AB} = \frac{\lambda}{2} \ep_{AB}.
\end{equation}

\noindent
However, there is still freedom if we define $\ep_{AB}$ to be a density instead of a tensor. We choose the weight of $\ep_{AB}$ such
that

\begin{equation}
\mathcal{L}_{\vv} \ep_{AB} = 0.
\end{equation}

\noindent
From definition \eqref{liespinor}:

\begin{align} 
\mathcal{L}_{\vv} \ep_{AB} = 0 &= \frac{\lambda}{2} \ep_{AB} + \mathrm{w}_{\ep} \partial_{m} \vv^{m} \ep_{AB} \nonumber \\
&= \left( \frac{1}{2} + 2 \mathrm{w}_{\ep} \right) \ep_{AB}
\end{align}

\noindent
we see this amounts choosing $\mathrm{w}_{\ep} = -1/4$. Consistency, however, requires
$\ep^{AB}$ to have weight $\mathrm{w}^{\ep} = +1/4$. 
Hence, given an arbitrary spinor $\mu^{A}$, in our conventions it is true that 

\begin{equation}
\mathcal{L}_{\vv} \mu_{A} = \ep_{AB} \mathcal{L}_{\vv} \mu^{B},
\label{weight}
\end{equation}

\noindent
which is equivalent to state that a spinor and its dual have the same conformal weight.
All considerations apply equally for dotted indices. 

%And with
%these choices, in the usual vector representation,
%the spacetime metric turns into a density. The transformation law for $\eta_{mn}$ changes to
%\begin{equation}
%\eta^{\prime}_{mn} \left( x^{\prime} \right) = \left\| \frac{\partial x^{\prime}}{\partial x} \right\|^{+2} \frac{\partial x^{a}}{\partial x^{\prime \, m }} \frac{\partial x^{b}}{\partial x^{\prime \, n }}
%\, \eta_{ab} (x) .
%\end{equation}

\subsection{Conformal invariance of Penrose action.}

In this section we will state the conformal invariance of the action 
\eqref{paction}. This in turn ensures the existence 
of a set of conformal symmetries in Fronsdal description.

Let us begin with dilatations. 
The higher-spin fields vary under it according to

\begin{subequations}
\begin{equation}
\delta_{\vv} \phi^{AB\cdots D} = \alpha x^{m} \partial_{m} \phi^{AB \cdots D} + 4 \alpha \mathrm{w}_{\phi} \phi^{AB \cdots D}
\end{equation}

\noindent
and

\begin{equation}
\delta_{\vv} a^{\da}_{\,\,\,\,\,\,B \cdots D} = \alpha x^{m} \partial_{m} a^{\da}_{\,\,\,\,\,\,B \cdots D} + 4 \alpha \mathrm{w}_{a} a^{\da}_{\,\,\,\,\,\,B \cdots D}.
\end{equation}
\end{subequations}

\noindent
These change the action by

\begin{align}
\delta_{\vv} S = \int \dd^{4} x &\left( \alpha x^{m} \partial_{m} \phi^{AB \cdots D} + 4 \alpha \mathrm{w}_{\phi} \phi^{AB \cdots D} \right) \partial_{A \da} a^{\da}_{\,\,\,\,\,\,B \cdots D} \nonumber \\
&+ \phi^{AB \cdots D} \partial_{A \da} \left( \alpha x^{m} \partial_{m} a^{\da}_{\,\,\,\,\,\,B \cdots D} + 4 \alpha \mathrm{w}_{a} a^{\da}_{\,\,\,\,\,\,B \cdots D} \right).
\end{align}

\noindent
After a few simplifications, we get

\begin{equation}
\delta_{\vv} S = \int \dd^{4} x \, \left\{ \alpha \left[ -3 + 4 \left( \mathrm{w}_{\phi} + \mathrm{w}_{a} \right) \right] \phi^{AB \cdots D} \partial_{A \da} a^{\da}_{\,\,\,\,\,\,B \cdots D} \right\},
\end{equation}

\noindent
which vanishes only when

\begin{equation}
\mathrm{w}_{\phi} + \mathrm{w}_{a} = \frac{3}{4}.
\label{dilations}
\end{equation}

\noindent
As we can see, dilatations are unable to fix completely the conformal weights. The remaining condition comes from the special conformal transformations.

Under special conformal transformations,
generated by

\begin{equation}
\vv^{m} = 2 \left( \rho . x \right) x^{m} - \left( x . x \right) \rho^{m}, 
\end{equation}

\noindent
the spin fields $\phi^{AB \cdots D}$ and $a^{\da}_{\,\,\,\,\,\, B \cdots D}$ vary according to

\begin{subequations}

\begin{equation}
\delta_{\vv} \phi^{AB \cdots D} = \vv^{m} \partial_{m} \phi^{AB \cdots D} + 2s \phi^{C (AB \cdots } f^{D)}_{\,\,\,\,\,\,C} + 8\mathrm{w}_{\phi} \left( \rho . x \right) \phi^{AB \cdots D},
\end{equation}

\noindent
and 

\begin{equation}
\delta_{\vv} a^{\da}_{\,\,\,\,\,\, B \cdots D} = \vv^{m} \partial_{m} a^{\da}_{\,\,\,\,\,\,B \cdots D} + \overline{f}^{\da}_{\,\,\,\,\,\,\dc} a^{\dc}_{\,\,\,\,\,\,B \cdots D} - \left(2s - 1 \right)
f^{C}_{\,\,\,\,\,\,(B} a^{\da}_{\,\,\,\,\,\, \cdots D)C} + 8 \mathrm{w}_{a} \left( \rho . x \right)  a^{\da}_{\,\,\,\,\,\,B \cdots D}.
\label{vouusar}
\end{equation}

\end{subequations}

\noindent
The action becomes

\begin{align}
\delta_{\vv} S &= \int \dd^{4} x \left( \vv^{m} \partial_{m} \phi^{AB \cdots D} \partial_{A \da} a^{\da}_{\,\,\,\,\,\,B \cdots D} + 2s \phi^{C (AB \cdots } f^{D)}_{\,\,\,\,\,\,C} 
\partial_{A \da} a^{\da}_{\,\,\,\,\,\, B \cdots D} + 8\mathrm{w}_{\phi} \left( \rho . x \right) \phi \partial a \right. \nonumber \\
&+ \left. \phi^{AB \cdots D} \partial_{A \da} \vv^{m} \partial_{m} a^{\da}_{\,\,\,\,\,\, B \cdots D} + \phi^{AB \cdots D} \vv^{m} \partial_{m} \partial_{A \da} a^{\da}_{\,\,\,\,\,\,B \cdots D} 
+\phi^{AB \cdots D} \partial_{A \da} \overline{f}^{\da}_{\,\,\,\,\,\,\dc} a^{\dc}_{\,\,\,\,\,\,B \cdots D} \right. \nonumber \\
&+ \left. \phi^{AB \cdots D} \overline{f}^{\da}_{\,\,\,\,\,\,\dc} \partial_{A \da}
a^{\dc}_{\,\,\,\,\,\,B \cdots D} 
- \left( 2s - 1 \right) \phi^{AB \cdots D} \partial_{A \da} f^{C}_{\,\,\,\,\,\, (B} a^{\da}_{\,\,\,\,\,\, \cdots D) C}  \right. \nonumber \\
&- \left. \left( 2s - 1 \right) \phi^{AB \cdots D} f^{C}_{\,\,\,\,\,\,(B} \partial_{|A \da|} a^{\da}_{\,\,\,\,\,\, \cdots D) C} + 8 \mathrm{w}_{a} \phi \rho a 
+ 8 \mathrm{w}_{a} \left( \rho . x \right) \phi \partial a \right).
\end{align}

\noindent
In the second line, we open $\partial_{a} \vv_{m}$ in its symmetric and anti-symmetric pieces and integrate by parts $\partial_{m}$ in $\partial_{A \da} \partial_{m} a^{\da}_{
\,\,\,\,\,\,B \cdots D}$. Then  we obtain

\begin{align}
\phi^{AB \cdots D} \partial_{A \da} \vv^{m} \partial_{m} a^{\da}_{\,\,\,\,\,\, B \cdots D} &=  \phi^{AB \cdots D} \partial_{(A \da} \vv_{m)} \partial^{m} a^{\da}_{\,\,\,\,\,\, B \cdots D}
+ \phi^{AB \cdots D} \partial_{[A \da} \vv_{m]} \partial^{m} a^{\da}_{\,\,\,\,\,\, B \cdots D} \nonumber \\
&= 2 \left( \rho . x \right) \phi \partial a + \phi^{AB \cdots D} f_{AM} \partial^{M}_{\,\,\,\,\,\, \da} a^{\da}_{\,\,\,\,\,\,B \cdots D} + \phi^{AB \cdots D} \overline{f}_{\da \dM} \partial_{A}^{\,\,\,\,\,\, \dM} a^{\da}_{\,\,\,\,\,\, B \cdots D}
\end{align}

\noindent
and 

\begin{align}
\phi^{AB \cdots D} \vv^{m} \partial_{m} \partial_{A \da} a^{\da}_{\,\,\,\,\,\,B \cdots D} &= - \vv^{m} \partial_{m} \phi^{AB \cdots D}  \partial_{A \da} a^{\da}_{\,\,\,\,\,\,B \cdots D}
- \partial_{m} \vv^{m}  \phi^{AB \cdots D}  \partial_{A \da} a^{\da}_{\,\,\,\,\,\,B \cdots D} \nonumber \\
&= - \vv^{m} \partial_{m} \phi^{AB \cdots D}  \partial_{A \da} a^{\da}_{\,\,\,\,\,\,B \cdots D} - 8 \left( \rho . x \right) \phi \partial a.
\end{align}

\noindent
When we substitute everything back into the action, the only remaining terms are

\begin{align}
\delta_{\vv} S = \int \dd^{4} x &\left\{ \left[ 8 \left( \mathrm{w}_{\phi} + \mathrm{w}_{a} \right) - 6 \right] ( \rho . x) \phi^{AB \cdots D} \partial_{A\da} a^{\da}_{\,\,\,\,\,\,B \cdots D} \right\}
+ \left( 8 \mathrm{w}_{a} - 3 \right) \phi^{AB \cdots D} \rho_{A \da} a^{\da}_{\,\,\,\,\,\,B \cdots D} \nonumber \\
&- (2s - 1) \phi^{AB \cdots D} \partial_{A\da} f^{C}_{\,\,\,\,\,\,(B} \, a^{\da}_{\,\,\,\,\,\, \cdots D)C}.
\end{align}

\noindent
We can use \eqref{mistake} so that

\begin{equation}
\partial_{A\da} f^{C}_{\,\,\,\,\,\,B} = - \rho_{\da}^{\,\,\,\,\,\,C} \ep_{AB} - \rho_{\da B} \delta_{A}^{\,\,\,\,\,\,C}.
\end{equation} 

\noindent
At the end, we get two relations involving the weights. They are

\begin{subequations}
\begin{equation}
8 \left( \mathrm{w}_{\phi} + \mathrm{w}_{a} \right) - 6 = 0
\label{trivia}
\end{equation}

\noindent
and

\begin{equation}
8\mathrm{w}_{a} + 2s - 4 = 0 .
\label{true}
\end{equation}
\end{subequations}

\noindent
If we use \eqref{dilations}, the first equation, \eqref{trivia}, is an identity. It gives no new information.
However, the second equation fixes the weight of the gauge field. Finally, we have

\begin{equation}
\mathrm{w}_{a} = \frac{2 - s}{4}
\end{equation}

\noindent
and

\begin{equation}
\mathrm{w}_{\phi} = \frac{s + 1}{4}.
\end{equation}

\noindent
The following table lists a few values for weights given different spin $s$ theories.

\begin{center}
\begin{tabular}{|c|c|c|}
\hline
\hspace{1.0 cm} & \hspace{.5cm} $\mathrm{w}_{\phi}$ \hspace{.5cm} & \hspace{.5cm} $\mathrm{w}_{a}$ \hspace{.5cm} \\
\hline
s = 0 & 1/4 & 1/2 \\
\hline
s= 1/2 & 3/8 & 3/8 \\
\hline 
s = 1 & 1/2 & 1/4 \\
\hline
s= 3/2 & 5/8 & 1/8 \\
\hline
s = 2 & 3/4 & 0 \\
\hline
s = 5/2 & 7/8 & -1/8 \\
\hline
\end{tabular}
\end{center}

\subsection{The structure of conformal transformations.}

Penrose theory is described by the set  $(\phi, a)$ while Fronsdal theory is
described by $h$. We have defined a map, which we name $H$, that takes one description into another:

\[
H: h_{m_{1} \cdots m_{s}} \longmapsto \left( \phi^{AB \cdots D} , a_{\da B \cdots D} \right).
\]

\noindent
It was shown that this map preserves phase space, i.e., it is a canonical transformation.

A map between symplectic structures also carries through symmetries of one description to another.
If a symplectic structure admits an action, then its symmetries must be also symmetries of the action.
Therefore it is natural to define a conformal transformation
of the form

\begin{equation}
\delta_{\vv} h_{m_{1} \cdots m_{s}} = H^{-1} \mathcal{L}_{\vv}\,  H \, \,h_{m_{1} \cdots m_{s}},
\label{notlie}
\end{equation}

\noindent
where $\vv$ is the conformal generator \eqref{conformalgenerator}.
%It must be a symmetry of Fronsdal action because it is a symmetry of Penrose action
%and both theories describe the same phase space.
It can act non-trivially; its action, as equation \eqref{notlie} shows,
is not obtained from standard Lie derivations.
Moreover, additional complications may appear due to $H^{-1}$, which involves
inverting derivatives, as \eqref{amaptoh} illustrates. 
For spins running from $s=1/2$ to $s=3/2$, it can be shown to agree with usual
conformal transformations obtained by change of coordinates. At spin $s=2$, however,
since Fronsdal theory \textit{is not} conformal invariant, our transformation exhibits
the non-local behaviour.

We can work out this case explicitly.
For special conformal transformations, if we plug the variation \eqref{vouusar} 
inside \eqref{map}, we obtain

\begin{equation}
\delta_{\vv} h_{(M \dM | N \dN)} \,\, = \,\, \mathcal{L}_{\vv} h_{(M \dM | N \dN )} + 2 \left( \rho . x \right) h_{(M \dM | N \dN )}
+ 6 \, \rho_{(\dM}^{\,\,\,\,\,\,\,\,\,\,E} \, a_{\dN) MNE} (h),
\end{equation}

\noindent
where $\mathcal{L}_{\vv}$, in this case, denotes the diffeomorphism Lie derivative
and $\rho$ is the special conformal parameter.
The last term shows the non-local behaviour since
it involves rewriting equation \eqref{vouusar} for $a_{\dM MNE}$ in terms of $h_{M\dM N \dN}$, giving inverse powers of $\partial_{a}$.
Notice that the conformal weight obtained from this expression, which reads $\mathrm{w} = + 1/4$, does not agree with 
the usual Fronsdal theory, which is dilatation invariant for $\mathrm{w} = - 1/4$ at
every spin\cite{Afr}.

These differences may appear problematic. They raise suspicion whether this 
transformation satisfies the conformal algebra or not. 
The simplest way to answer this question is to notice that \eqref{notlie} is a conjugation;
therefore, if $H$ is well-defined, they must satisfy the same algebra
of the vector field $\vv$ in question.

\section{Conclusions.}

We have defined an action for Penrose theory and constructed its symplectic structure.
This action appears to be simpler than the usual one obtained by Fronsdal. Moreover,
it depends only on the epsilon symbol, being possible to examine how it should extend
to curved spaces. It would be interesting to see how it compares with 
Vasiliev theory in $AdS_{4}$.

In this paper, we addressed a different question. We showed that both theories describe 
the same classical phase space. It, in turn, leads us to conjecture a set of
non-trivial conformal symmetries for the Fronsdal higher-spin field $h_{m_{1} \cdots m_{s}}$.
These are not generated by usual coordinate changes, although to lower spins --
those which run from $1/2$ to $3/2$ -- it is possible to show that both symmetries agree.
The non-local behaviour appears only at spin $2$.
This consideration raises the question of how these new symmetries would compare with
Segal's formulation of conformal higher-spin theories \cite{segal}.

\vspace{.5cm}
\noindent
\textbf{Acknowledgements.} I am indebted to Alexei Rosly and Andrei Mikhailov for 
suggesting this problem and for numerous helpful discussions.
I also would like to thank Arkady Tseytlin for reading the manuscript and 
for useful suggestions. This work was supported by CAPES grant 33015015001P7 and by
FAPESP grant 2014/18634-9.

\appendix
\numberwithin{equation}{section}
\section{A mini-introduction to the geometry of classical mechanics.} \label{appendix}

This appendix explains the terminology used in this work.
We briefly review basic aspects of the geometry of classical mechanics.

The classical phase space, $M$, is the set of all classical trajectories.
This space is naturally an infinite-dimensional symplectic manifold, that is,
a pair $(M, \Omega)$ consisting a smooth manifold $M$ and a non-degenerate closed 2-form 
$\Omega$ called symplectic structure. 

Let us explain how to obtain the symplectic structure from the action.
Fix

\begin{equation}
S[\phi(x)] = \int \mathcal{L} \left( \phi(x) , \partial_{m} \phi (x) \right) \dd^{4} x
\end{equation}

\noindent
for a given field $\phi(x)$
and let the classical configuration be denoted $\phi_{cl} (x)$.
Then, under arbitrary infinitesimal changes in field configuration, for example $\delta \phi (x)$, the action changes around the classical
path according to 

\begin{equation}
S\left[\phi_{cl}(x) -  \delta \phi(x)\right] - S \left[ \phi_{cl}(x)  \right] = \int_{C}  \frac{\partial \mathcal{L}}{\partial (\partial_{m} \phi)} n_{m} \dd^{3} x \wedge \delta \phi (x),
\end{equation}

\noindent
where $n_{m} \dd^{3} x$ defines a $3$-form in Minskowski space to be integrated over $C$, a $3$-dimensional closed surface.

\vspace{.5cm}
\textbf{Remark.}
Here we have the de Rham complex with exterior derivative $\dd$
and the variational complex with differentiation $\delta$; the previous variation $\delta \phi (x)$ may be interpreted as a differential
form on the space of field configurations. When dealing with $\dd$ and $\delta$, we will use the following rules:

\[
\dd \delta = - \delta \dd \quad \text{and} \quad \delta \phi(x) \wedge \dd x^{m} = - \dd x^{m} \wedge \delta \phi(x).
\]

\vspace{.5cm}
The variation $\delta \phi$  descends to the phase space once we take it to satisfy the equations of motion.
One then can consider formally the symplectic structure to be

\begin{equation}
\Omega = \int_{C} \delta \left( \frac{\partial \mathcal{L}}{\partial (\partial_{m} \phi)} \right)  \wedge \delta \phi (x) \wedge n_{m} \dd^{3} x ,
\label{symp}
\end{equation}

\noindent
since it defines a closed $2$-form on phase space.
Such differential form is also independent of $C$. To see this, consider for example two contours, 
$C_{1}$ and $C_{2}$. And let $\Omega_{1}$ and $\Omega_{2}$ represent the respective symplectic structures.
We want to show that

\begin{equation}
\Omega_{1} - \Omega_{2} = 0
\end{equation}

\noindent
in $M$. 
Define $\Sigma$ to be the $4$-dimensional surface whose boundary is $C_{1} - C_{2}$, then by Stokes' theorem

\begin{align}
\Omega_{1} - \Omega_{2} &= \int_{\Sigma} \partial_{m} \left( \delta \frac{\partial \mathcal{L}}{\partial (\partial_{m} \phi)}  \wedge \delta \phi (x) \right) \wedge  \dd^{4} x \nonumber \\
&= \int_{\Sigma} \delta  \left( \partial_{m} \frac{\partial \mathcal{L}}{\partial (\partial_{m} \phi)} \right) \wedge \delta \phi (x) \wedge  \dd^{4} x + \delta \frac{\partial \mathcal{L}}{\partial (\partial_{m} \phi)} 
\wedge \delta \partial_{m} \phi (x) \wedge \dd^{4} x \nonumber \\
& = \int_{\Sigma} \delta \frac{ \partial \mathcal{L}}{\partial \phi} \wedge \delta \phi (x) \wedge \dd^{4} x + \delta \frac{\partial \mathcal{L}}{\partial (\partial_{m} \phi)} 
\wedge \delta \partial_{m} \phi (x) \wedge \dd^{4} x \nonumber \\
& = \int_{\Sigma} \delta^{2} \mathcal{L} \wedge \dd^{4} x = 0
\end{align}

\noindent
with the help of Euler-Lagrange equations.
In this computation, and in all of those which involve a symplectic structure, we stress that we are free to use the equations of motion
because we are in phase space.

%Contact with usual classical mechanics is made as follows. 
%A Hamiltonian vector field $V_{h}$ defines a Hamiltonian function $h$ if
%\begin{equation}
%\iota_{V_{h}} \Omega = \delta h.
%\end{equation}
%\noindent
%Then two Hamiltonian vector fields $V_{h}$ and $U_{f}$ define the poisson bracket $\{h, f\}$ 
%to be
%\begin{equation}
%\left\{ h, f \right\} = \iota_{U_{f}} \iota_{V_{h}} \Omega.
%\end{equation}

%For example, if we take \eqref{symp}
%then we get 
%\begin{equation}
%\theta \left( \cdot \,, \cdot \right) = \int_{C} \left( \frac{\delta \, \cdot}{\delta \pi_{m}(x) } \frac{ \delta \, \cdot}{\delta \phi(x)} 
%- \frac{\delta \, \cdot}{\delta \phi (x)} \frac{\delta  \, \cdot}{\delta \pi_{m} (x)} \right) n_{m} \dd^{3} x,
%\end{equation}
%\noindent
%where we have defined 
%\[
%\pi_{m} (x) = \frac{\partial \mathcal{L}}{\partial ( \partial_{m} \phi) }.
%\]

In classical mechanics, a symplectic structure defines a Poisson bracket.
For example, one can consider, in a local basis, a 
bivector which is the inverse matrix of the symplectic form.
This bivector, by definition, maps functions into functions and 
satisfies the Jacobi identity -- a consequence of the closeness of $\Omega$.  

\vspace{.5cm}
\noindent
\textbf{Examples.}

\vspace{.2cm}
\noindent
\textsl{Spin} $s = 0$. The action is

\begin{equation}
S = \int \dd \phi \wedge \star \dd \phi
\end{equation}

\noindent
and the symplectic structure obtained is

\begin{equation}
\Omega = \int_{C} \delta \phi \wedge \star \dd \delta \phi.
\end{equation}

\noindent
One can choose the surface $C$ to be $t = \mathtt{constant}$ and the symplectic structure 
turns into

\begin{equation}
\Omega = \int \delta \phi \wedge \delta \dot{\phi} \wedge \dd^{\,3} \mathbf{x}.
\end{equation}

\noindent
Its inverse gives rise to the Poisson Bracket

\begin{equation}
\left\{ F, G \right\} = \int \left( \frac{\delta F}{\delta \phi(\mathbf{x})} \frac{\delta G}{\delta \dot{\phi}(\mathbf{x})} 
- \frac{\delta G}{\delta \phi(\mathbf{x})} \frac{\delta F}{\delta \dot{\phi}(\mathbf{x})} \right) \dd^{3} \mathbf{x}.
\end{equation}

\vspace{.2cm}
\noindent
\textsl{Spin} $s=1/2$. The action is

\begin{equation}
S = \int \overline{\psi}^{\da} \partial_{\da A} \psi^{A}
\end{equation}

\noindent
and the symplectic structure obtained is

\begin{equation}
\Omega = \int_{C} \delta \overline{\psi}^{\da} \wedge \delta \psi^{A} \wedge \dd^{3} x_{\da A}.
\end{equation}

\noindent
One can choose the surface $C$ to be $t= \mathtt{constant}$ and the symplectic structure turns
into

\begin{equation}
\Omega = \int \delta \overline{\psi}^{\da} \wedge \psi^{A} \wedge  \sigma^{0}_{A \da} \dd^{3} \mathbf{x}.
\end{equation}

\noindent
Its inverse gives rise to the Poisson Bracket

\begin{equation}
\left\{ F, G \right\} = \int \left( \frac{\delta F}{\delta \overline{\psi}^{\da}(\mathbf{x})} \frac{\delta G}{\delta \psi^{A}(\mathbf{x})}
- \frac{\delta G}{\delta \overline{\psi}^{\da}(\mathbf{x})} \frac{\delta F}{\delta \psi^{A}(\mathbf{x})} \right) \overline{\sigma}^{0 A \da} \, \dd^{3} \mathbf{x}.
\end{equation}

\vspace{.5cm}
The set of transformations that preserve the symplectic structure will also 
preserve the Poisson bivector. These are usually called canonical transformations.

\end{document}